\documentclass{vgtc}                          

\ifpdf
  \pdfoutput=1\relax                   
  \usepackage{graphicx}                
  \DeclareGraphicsExtensions{.pdf,.png,.jpg,.jpeg} 
\else
  \usepackage{graphicx}                
  \DeclareGraphicsExtensions{.eps}     
\fi%

\graphicspath{{figures/}{pictures/}{images/}{./}} 

\usepackage{microtype}                 
\PassOptionsToPackage{warn}{textcomp}  
\usepackage{textcomp}                  
\usepackage{mathptmx}                  
\usepackage{times}                     
\usepackage{cite}                      
\usepackage{tabu}                      
\usepackage{booktabs}                  

\onlineid{0}

\vgtccategory{Research}

\vgtcinsertpkg

\title{Using Resource-Rational Analysis to Understand Cognitive Biases in Interactive Data Visualizations}

\author{Ryan Wesslen\thanks{e-mail: rwesslen@uncc.edu}\\ %
        \scriptsize UNC Charlotte %
\and Douglas Markant\thanks{e-mail: dmarkant@uncc.edu}\\ %
     \scriptsize UNC Charlotte %
\and Alireza Karduni\thanks{e-mail: akarduni@uncc.edu}\\ %
     \scriptsize UNC Charlotte
\and Wenwen Dou\thanks{e-mail: wdou1@uncc.edu}\\ %
     \scriptsize UNC Charlotte
     }

\abstract{Cognitive biases are systematic errors in judgment. Researchers in data visualizations have explored whether cognitive biases transfer to decision-making tasks with interactive data visualizations. At the same time, cognitive scientists have reinterpreted cognitive biases as the product of \emph{resource-rational} strategies under finite time and computational costs. In this paper, we argue for the integration of resource-rational analysis through constrained Bayesian cognitive modeling to understand cognitive biases in data visualizations. The benefit would be a more realistic ``bounded rationality'' representation of data visualization users and provides a research roadmap for studying cognitive biases in data visualizations through a feedback loop between future experiments and theory.  
} 

\CCScatlist{
  \CCScatTwelve{Human-centered computing}{Visu\-al\-iza\-tion}{Visualization theory, concepts and paradigms}{}
}

\begin{document}


\firstsection{Introduction}

\maketitle

Recently data visualization researchers have investigated whether cognitive biases transfer to decision-making with interactive visual interfaces and explored strategies to mitigate them \cite{wall2017warning,dimara2018mitigating,dimara2018task,cho2017anchoring,wesslen2019investigating}. A large portion of this work involves collecting and analyzing empirical evidence on the effect of different cognitive biases through user experiments. These studies generally are motivated by the classical psychological approach to cognitive biases, i.e., the ``heuristic and biases'' framework \cite{streeb2018biases} introduced by Tversky and Kahneman \cite{tversky1973availability,tversky1974judgment}. While these studies provide great value to the visualization community on illuminating the effect of cognitive biases on visual analysis tasks, they do not include quantitative cognitive models that yield explicit testable hypotheses predicting users' behavior under different experiment conditions. Moreover, they tend to ignore critiques that heuristics which lead to biased judgments actually reflect the rational use of limited cognitive resources \cite{gigerenzer2011heuristic}.


Adopting cognitive modeling for data visualization research can provide many opportunities to accelerate innovation, improve validity, and facilitate replication efforts \cite{padilla2018case}. Early examples to understand cognitive processes while using data visualizations consider sensemaking approaches \cite{green2009building} or visual attention coupled with decision-making \cite{padilla2018decision}. 
However, a drawback of past Vis cognitive frameworks \cite{green2009building,patterson2014human,padilla2018decision} is they are typically descriptive or ``process'' diagrams.
As such, they lack the level of detail necessary to make detailed quantitative predictions or to generate strong hypotheses about behavior \cite{farrell2010ComputationalModelsAids}.
This complicates efforts to predict when cognitive biases will impact how people interpret and make decisions from visualizations.

One area of opportunity is Bayesian cognitive modeling for data visualizations \cite{wu2017towards,kim2019bayesian,karduni2020beliefs}.
These models rest on the claim that people reason under uncertainty in accordance with the principles of Bayesian inference \cite{griffiths2015rational}.
This approach is appealing because it provides a normative framework for how people should reason and make decisions from information under uncertainty.
However, in practice people may behave in ways that are inconsistent with the predictions of Bayesian models, often due to well-known limitations in cognitive capacity, including constraints in time, which are common in many visualization studies and tasks.
Existing applications of Bayesian cognitive modeling to information visualization have not yet acknowledged these limitations like bounded rationality \cite{simon1955behavioral,simon1956rational}.

Specifically, we'd like to note gaps and disconnect among current efforts in studying the effect of cognitive biases in interactive data visualizations. A promising approach from cognitive science is the \textbf{resource-rational analysis} of cognitive biases as a way to understand rational trade-offs between judgment accuracy and the mind's limited resources, including fixed time \cite{griffiths2015rational,lieder2018anchoring,lieder2020resource}.  
In this paper, we argue that resource-rational analysis can provide a framework for many cognitive biases in data visualizations while providing a quantitative theoretical framework, or ``research roadmap,'' that enables a feedback  loop to add realism through further constraints. 
Such a roadmap may not only better identify cognitive biases' effects in data visualization decision-making but may also provide a means for mitigating these biases before they occur.

\section{Cognitive Bias in Interactive Data Visualizations}


Cognitive biases are systematic errors (or deviations) in judgment \cite{tversky1974judgment,lieder2020resource}. They have been  studied by cognitive psychologists and social scientists to understand how and why individuals sometimes make consistent errors in decision-making. Recently, data visualization researchers have explored the role of cognitive biases transfer to data visualization decision-making \cite{wall2017warning,dimara2018task,valdez2018studying} and, if such biases can be identified, how these findings could inform the design of visualizations systems that can debias or mitigate such effects \cite{pohl2014sensemaking,ellis2018so,dimara2018mitigating,parsons2018promoting}. If a well-designed system can help users to find the right explore-exploit mix \cite{hills2015exploration}, ideally such a system would safeguard against possible forking path problems \cite{pugarden, zgraggen2018investigating} and mitigate systematic errors and enable better decision-making.



Data visualization research in cognitive biases tend to focus on either empirical studies or frameworks with little interaction between them.
Empirical studies try to demonstrate evidence of traditional cognitive biases through data visualization user studies, typically analyzing user's interaction behaviors or decisions \cite{dimara2017attraction,dimara2018mitigating,cho2017anchoring,wesslen2019investigating,valdez2018priming,karduni2018icwsm}. Alternatively, general descriptive frameworks (like taxonomies) have been introduced for cognitive biases \cite{dimara2018task,wall2017warning}; however, these tend to broadly cover many human biases \cite{valdez2018studying,wall2017four} and are limited in their ability to provide testable predictions for empirical studies.

Cognitive science has a long history of studying visualization cognition as a subset of visuospatial reasoning, in how individuals derive meaning from visual (external) representations \cite{tversky2005visuospatial}. Typically, these models either focused on perception and/or prior knowledge \cite{padilla2018decision}. More recently, data visualization researchers have integrated similar ideas to understand visualization cognitive processes through insight-based approaches \cite{green2009building} and top-down modeling \cite{liu2010mental,patterson2014human}. 
However, past visualization cognitive models tend to be based on verbal ``process'' diagrams, and are not quantitative models that yield explicit testable hypotheses. Without such quantitative predictions, implications of the models can be vague, difficult to simulate, and even more difficult to test and refine. 

Bayesian cognitive modeling is a promising approach to studying cognitive biases \cite{wu2017towards}. Building on work in cognitive science, Wu \textit{et al.} first argued that Bayesian cognitive modeling provides a means to model many irrational behaviors like cognitive biases in a ``principled'' way. Building from their work, Kim \textit{et al.} \cite{kim2019bayesian} and Karduni \textit{et al.} \cite{karduni2020beliefs} have provided further extensions on studying Bayesian cognitive modeling for data visualizations. In particular, by eliciting each user's prior belief about an uncertain relationship, these studies have used Bayesian models to predict how people should update those beliefs in response to data visualizations. Although these two studies provide novel elicitation methods with Bayesian cognitive models in data visualization, they do not directly connect such approaches with experiment designs to identify cognitive biases. Moreover, they do not incorporate realistic constraints on users (e.g., time or memory limits) in their modeling or experiment. This is where \textbf{resource-rational analysis} may remedy these shortcomings.



\section{Resource-Rational Analysis}

 Classical approaches to understand rationality \cite{zouboulakis2014varieties,nozick1994nature} assume individuals incorporate utility theory \cite{vonneuman1944} to maximize their expected utility. Simon \cite{simon1955behavioral,simon1956rational} challenged this notion with \textbf{bounded rationality}, the idea that rational decisions must be framed in the context of the environment and one's limited cognitive resources. 
 Whereas normative rational models exist on Marr's computational level \cite{marr1982vision} (i.e., on the structure of the problem), bounded rationality connects Marr's computational level and the algorithmic level (e.g., representation and transformation) as human cognition involves making approximations from a normative rational model \cite{griffiths2015rational,lieder2018resource}. The problem is studying each level separately is insufficient to explain the underlying mechanisms in human intelligence \cite{lieder2018resource}. 

\begin{figure}
\includegraphics[width=0.5\textwidth]{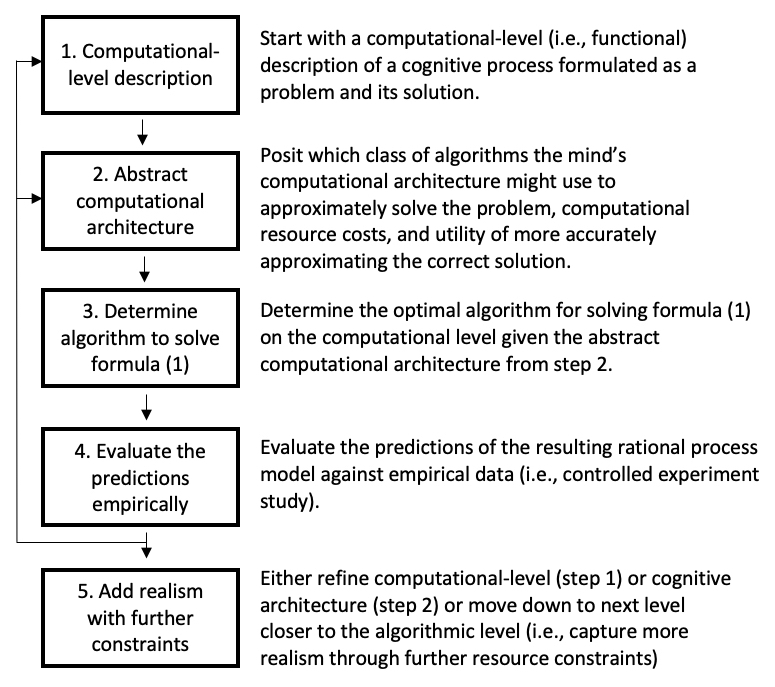}
\caption{Flow diagram of five steps of the resource-rational analysis process adapted from Lieder and Griffiths \cite{lieder2018resource}.}
\label{fig:resource-rational-flow}
\end{figure}

To address this problem, Lieder and Griffiths \cite{lieder2018anchoring,lieder2018resource,lieder2020resource} introduce \textbf{resource-rational analysis} as rational models that bridge the idealized, unbounded computational level to a more realistic, highly resource-constrained algorithmic level. As an iterative process, a rational model can be modified over time to move closer towards a more realistic model of individuals' true cognitive resources and processes. Figure \ref{fig:resource-rational-flow} outlines the five steps in resource-rational analysis. Like  other rational theories, resource-rational theory posits that there exists some optimal solution yielded by the rules of expected utility theory, Bayesian inference, and standard rules of logic (Step 1 in Fig. 1). However, bounded rationality limits the space of feasible decisions that are possible given the cognitive constraints which lead to approximate models of rationality (Step 2). Instead, resource rationality is the optimal algorithm under this constraint (Step 3) which then yield testable predictions (Step 4). In this way, resource-rational analysis reinterprets cognitive biases as an optimal (rational) tradeoff between external task demands and internal cognitive constraints (e.g., cost of error in judgment vs. time cost to reduce this error) \cite{lieder2018empirical}. This rational interpretation reconciles with Gigerenzer's criticism of cognitive biases as irrational use of heuristics as rational \cite{gigerenzer2011heuristic}.




\subsection{Example: Anchoring Bias}

One popular cognitive bias that has been studied in multiple visualization experiments is anchoring bias \cite{cho2017anchoring,wesslen2019investigating,wall2019formative,valdez2018priming}. Anchoring bias is the tendency for an initial piece of information, relevant or not, to effect a decision-making process \cite{tversky1974judgment}. Typically, this is followed by an adjustment in response to new information which falls short of the normative judgment (the anchoring-and-adjustment effect). Anchoring-and-adjustment approach posits a two step process \cite{epley2006anchoring}. In the first step, a person will develop their estimate, or anchor, of an open-ended question. In the second step, the person will adjust her estimate as new information is processed. Error occurs when she does fails to make a sufficient adjustment to the correct answer. 

Lieder and Griffiths \cite{lieder2018anchoring} examined anchoring bias through the lens of resource-rational analysis. Following Figure 1, they formulate the problem through Bayesian decision theory for numerical estimation, the classical task  associated with anchoring-and-adjustment \cite{tversky1974judgment,epley2006anchoring}.  They assume that the mind approximates Bayesian inference through \textbf{sampling algorithms}, which represent probabilistic beliefs through a small number of randomly selected hypotheses proportional to their actual prevalence \cite{vul2010sampling}. More specifically, they posit that sampling occurs through \textbf{Markov Chain Monte Carlo} (\textbf{MCMC}), a popular algorithm in statistics and artificial intelligence.



The advantage of this approach is that it provides testable predictions that can be considered empirically through controlled experimentation (Step 4 in Fig. 1). The model predicts that scenarios in which there are high time costs and no error costs, result in the highest degree of anchoring bias as participants have a much higher cost for each adjustment but less concern for accuracy (or error). Therefore, in such situations participants will tend to have more bias (absolute distance). This occurs as participants have zero adjustments and favor their anchor (provided or self-generated) as time costs are critical. To test this model, Lieder \textit{et al.} \cite{lieder2018empirical} developed an empirical experiment on MTurk for estimating bus arrival under four different scenarios. They find strong evidence for resource rationality adjustment as the degree of anchoring bias varied based on different time and error costs. Moreover, they find that incentives can be effective at reducing anchoring bias even with self-generated and provided anchors, contrary to Epley and Gilovich \cite{epley2006anchoring}.


\section{Future Work and Conclusion}

Resource-rational analysis could be beneficial in data visualization studies in which users are faced with meaningful cost-benefit tradeoffs in interpreting the visualization. In other words, experiments where additional effort leads to a more accurate decisions from the data. This would especially be the case for system in which sampling occurs over time, either directly sampling information from a display, or sampling alternative states/outcomes in the user's mental model. In the context of visualizing hurricane paths \cite{ruginski2016non}, users might at first overweight the risks of salient negative outcomes (e.g., a direct hit on New Orleans), but with more time (or different type of visualization?) arrive at a better calibrated estimate of the risk.


\bibliographystyle{abbrv-doi}

\bibliography{template.bbl}
\end{document}